# Affine Frequency Division Multiplexing (AFDM) for 6G: Properties, Features, and Challenges

Hyeon Seok Rou, Kuranage Roche Rayan Ranasinghe, Vincent Savaux,
Giuseppe Thadeu Freitas de Abreu, David González G., and Christos Masouros

## Abstract

Affine frequency division multiplexing (AFDM) is an emerging waveform candidate for future sixth-generation (6G) systems offering a range of promising features, such as enhanced robustness in heterogeneous and high-mobility environments, as well as inherent suitability for integrated sensing and communications (ISAC) applications. In addition, unlike other candidates such as orthogonal time-frequency space (OTFS) modulation, AFDM provides several unique advantages that strengthen its relevance to practical deployment and standardization in 6G. Notably, as a natural generalization of orthogonal frequency division multiplexing (OFDM), strong backward compatibility with existing conventional systems is guaranteed, while also offering novel possibilities in waveform design, for example to enable physical-layer security through its inherent chirp parametrization. In all, this article provides an overview of AFDM, emphasizing its suitability as a candidate waveform for 6G standardization. First, we provide a concise introduction to the fundamental properties and unique characteristics of AFDM, followed by highlights of its advantageous features, and finally a discussion of its potential and challenges in 6G standardization efforts and representative requirements.

## Introduction

The telecommunications community, including academia and industry, is working vigorously in the development of sixth-generation (6G) wireless systems expected to be standardized by 2030 [1]. The objective for 6G is to deliver transformative capabilities, including peak data rates up to 1 Terabits-per-second (Tbps), ultra-low latency below 1 millisecond (ms), and massive connectivity supporting up to $10^7$ devices per km$^2$. These ambitious goals are driven by the six well-known IMT-2030 usage scenarios [2], including integrated sensing and communications (ISAC), ubiquitous connectivity, and immersive communication, which are complemented by selected use cases such as Internet-of-Things (IoT), vehicle-to-everything (V2X), space-air-ground integrated networks (SAGINs), and digital twin networks (DTNs) [3].

Enabling such advanced applications and satisfying such stringent requirements demand, however, wireless systems to operate in increasingly higher frequency bands, such as the millimeter wave (mmWave) and Terahertz (THz) spectra, subject to multipath and high-mobility. This gives rise to doubly-dispersive channels that exhibit both time- and frequency-selectivity, leading to severe deterioration of received signals and, consequently, substantial challenges for communications systems employing conventional waveforms [4]. In particular, the orthogonal frequency division multiplexing (OFDM) waveform, widely used in fourth generation (4G) and fifth generation (5G) systems, *struggles* in such environments due to severe inter-symbol interference (ISI) and inter-carrier interference (ICI) caused by multipath delays and especially Doppler shifts, which disrupts OFDM subcarrier orthogonality.

Furthermore, 6G is anticipated to bring a paradigm shift towards a *native*, and hence, optimized, support of ISAC [1], enabling a wide range of multifunctional services, real-time situational awareness, improved spectrum utilization, and high-resolution localization [5]. These features are not easily available through conventional waveforms, which are susceptible to Doppler interference and prevent accurate velocity estimation, thus motivating the design of novel waveforms that can offer strong Doppler robustness and sensing capabilities.

Given the above, innovative waveform design for 6G has gained great interest in recent years. In particular, in order to address the aforementioned challenges, researchers have explored various alternatives to maintain subcarrier orthogonality in doubly-dispersive conditions. Examples of waveforms that meet such requirements are: orthogonal time-frequency space (OTFS), whereby data is modulated in the delay-Doppler (as opposed to time-frequency) domain, improving robustness against Doppler fading [6]; and orthogonal chirp division multiplexing (OCDM),

*Hyeon Seok Rou (corresponding author), Kuranage Roche Rayan Ranasinghe, and Giuseppe Thadeu Freitas de Abreu are with the School of Computer Science and Engineering, Constructor University, 28759 Bremen, Germany; Vincent Savaux is with the Institute of Research and Technology b-com, 35510 Cesson-Sévigné, France; David González G. is with Wireless Communications Technologies, AUMOVIO Germany GmbH, 60488 Frankfurt am Main, Germany; Christos Masouros is with the Department of Electrical and Electronic Engineering, University College London, WC1E 6BT London, U.K.*

which leverages chirp-shaped subcarriers for enhanced performance in dispersive channels [7].

Another recent proposal is affine frequency division multiplexing (AFDM), which has emerged as a promising candidate for 6G [8], whereby data based is modulated via the affine Fourier transform (AFT), an affine generalization of the conventional linear Fourier transform (FT) employed in OFDM. By leveraging two tunable parameters that control the *twistedness* of the affine time-frequency domain, AFDM is capable of adapting to the specific characteristics of doubly-dispersive channels, ensuring optimal diversity gain – i.e., exploit multiple independently fading paths with varying time-frequency selectivity for improved reliability. In addition, these tunable parameters facilitate a wide range of waveform design opportunities, supporting inherent advanced functionalities, such as index modulation (IM), peak-to-average power ratio (PAPR) reduction, and physical-layer security.

Owing to its inherent delay-Doppler localization, directly related to the range and velocity sensing capabilties of the waveform, recent studies have also identified AFDM as a highly effective waveform for ISAC which can well address the communications-sensing trade-off [9]. In such, AFDM offers two fundamental ISAC approaches: a radar-like correlation approach, where the favorable ambiguity function (AF) of AFDM in both delay and Doppler domains enables high range-velocity resolution [10], and another approach based on the deterministic structure of the AFDM effective channel, whose delay-Doppler representation allows channel estimation to be equivalent to the radar parameter estimation, or vice versa [11]. Both approaches have demonstrated strong performances and highlight the potential of AFDM-based ISAC in 6G.

Amidst the research, there are also ongoing discussions on candidate waveforms for 6G, where a key point of debate among industry, academia, and standardization bodies, such as the 3rd Generation Partnership Project (3GPP) and European Telecommunications Standards Institute (ETSI), continues to be whether 6G necessitates entirely new waveform paradigms or refinements of existing ones. In other words, rather than replacing legacy waveforms like OFDM, many advocate for expanding the waveform toolbox to enable flexible, context-aware operation, highlighting the need for candidates that balance performance, compatibility, and deployability.

Indeed, as of October 2025, several companies within the 3GPP are actively proposing new waveform candidates, among which AFDM has been highlighted as a promising option [12]. This interest stems from its inherent compatibility with legacy waveforms such as CP-OFDM and DFT-s-OFDM, which enables seamless integration and coexistence with existing 4G/5G infrastructure, as well as from its channel adaptability – an essential feature for the diverse use cases and deployment scenarios envisioned for 6G [13].

All in all, the above establishes AFDM as a strong candidate waveform for 6G. Accordingly, this article offers a comprehensive and standardization-oriented study of AFDM, examining its key properties, advantages, and challenges specifically in relation to the 6G requirements and deployment considerations. This perspective well complements other studies, for example which focus on AFDM's waveform characteristics and flexibility across various channel and application scenarios [13], in all providing a holistic view of the waveform's potential and practical relevance.

The article is structured as follows. The "Fundamentals and Key Properties" section details the fundamentals and key properties of the AFDM waveform; the "AFDM: Features and Applications" section examines eight of the advantageous features of AFDM and evaluates their implications to relevant 6G applications; the "Standardization Considerations for 6G: Challenges, Opportunities, and Pathways" section then addresses another set of eight points on opportunities, challenges, and pathways towards standardization of AFDM in future 6G systems; and finally, the "Conclusion and Outlook" section concludes with a summary and prospects.

## Fundamentals and Key Properties

In this section, we briefly introduce the fundamentals of the AFDM signal modulation, highlighting the inherent properties of the waveform, providing a background towards the consequent discussions to be elaborated in the following sections.[1]

### Signal Generation and Detection

At each transmission instance, a block of $N$ information symbols is mapped onto $N$ subcarriers through the inverse discrete affine Fourier transform (IDAFT), which serves as the modulation kernel of AFDM. Unlike the conventional notion of static, sinusoidal subcarriers in OFDM (i.e., puretone), which maintain orthogonality in the frequency domain, AFDM employs *chirp*-based subcarriers that sweep across the frequency bandwidth over time, as illustrated in the subcarrier visualizations of Fig. 1. Note that despite their time-varying nature, the $N$ chirp subcarriers maintain mutual orthogonality across both time and frequency domains, forming a robust basis well-suited for time-varying propagation environments.

Furthermore, the frequency *sweep rate* of the chirp subcarriers in AFDM is fully configurable via the two tunable chirp parameters inherent to the IDAFT (often denoted by $c_1$ and $c_2$), namely by $c_1$. As will be further elaborated upon later, this unique design flexibility through the inclusion of tunable chirp parameters - not present in contending waveforms - is a key feature of AFDM, which ensures orthogonality and full diversity in doubly-dispersive channels, in addition to enabling a plethora of waveform optimization and applications.

Trivially, the IDAFT of AFDM parallels the role of the inverse discrete Fourier transform (IDFT) modulator in OFDM, and also stands in contrast to other chirp-subcarrier based waveforms such as OCDM leveraging the inverse discrete Fresnel transform (IDFnT) for modulation, in which the freqeuncy sweep rate is fixed and not configurable.

Considering the implementation of the modulator, the core IDAFT can be efficiently be decomposed into a chirp-domain pre-processing operation, followed by a standard IDFT, and a chirp-domain post-processing stage, as visualized in Fig. 1. This structure enables efficient implementation using the conventional fast Fourier transform (FFT) technology core to the 4G and 5G architectures. Furthermore, recent works have even demonstrated that both chirp pre-/post-processing steps be implemented using only discrete Fourier transform (DFT)/IDFT operations, consolidating the overall computational efficiency and compatibility of the AFDM modulator.

[1] Given the breadth of the audience and the nature of the IEEE Magazine, we present an illustrative and intuitive overview of the AFDM fundamentals, omitting detailed derivations. Readers interested in the full theoretical formulation and analysis of AFDM are encouraged to consult [4], [8].

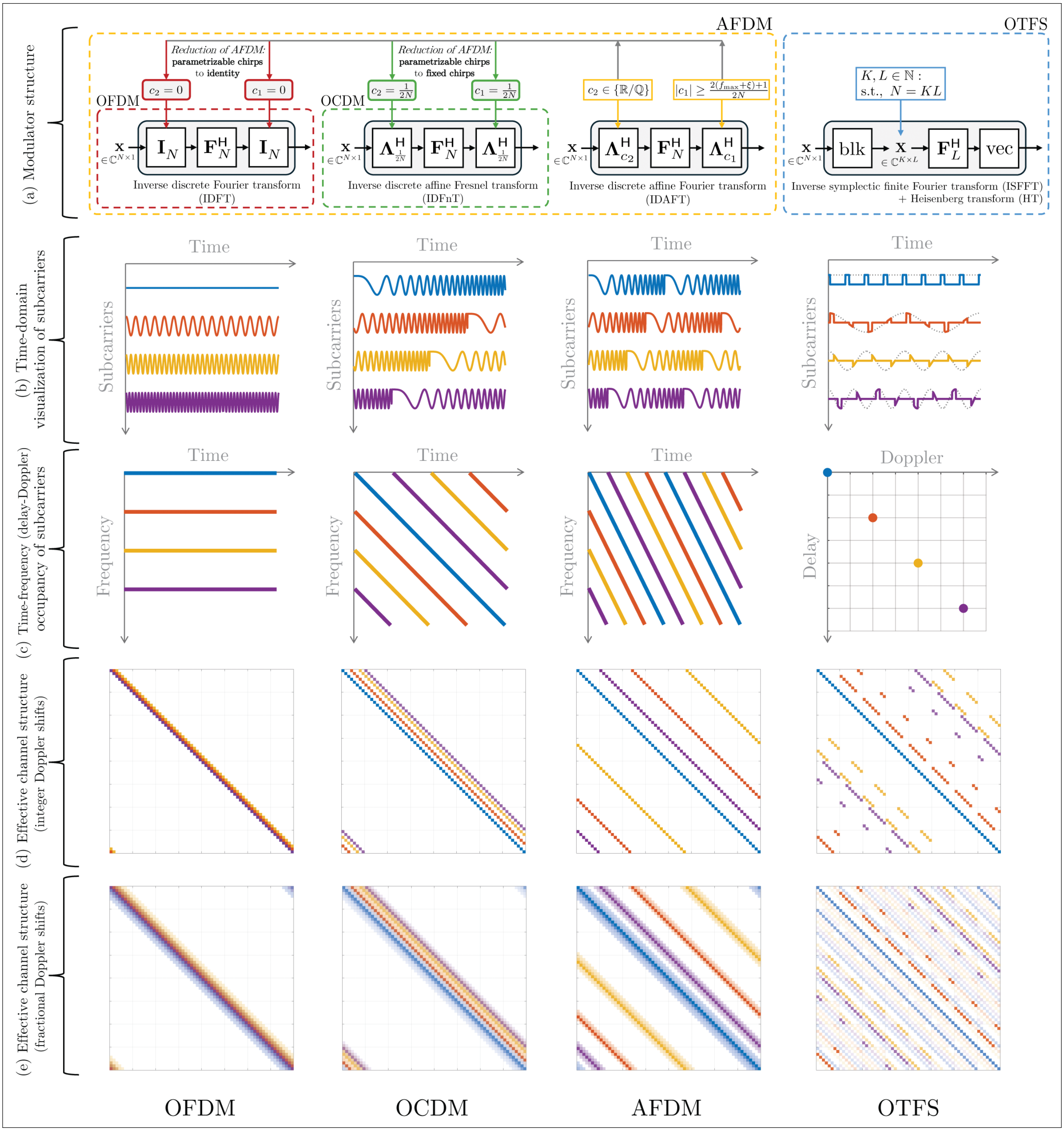

**FIGURE 1.** Illustrative comparison of OFDM, OCDM, AFDM, and OTFS waveforms in terms of: a) the modulator structure in terms of the core Fourier transform and the relationship of the chirp frequencies $c_1$ and $c_2$, which shows that AFDM generalizes OFDM and OCDM, b) the time-domain subcarriers, c) the time-frequency (delay-Doppler) occupancy of the subcarriers, d) effective channel structure with four arbitrary integer delay-Doppler taps, and e) the same channels of d) with fractional Doppler shifts (i.e., normalized Doppler frequencies do not align perfectly with the discrete integer grid used in delay-Doppler signal processing), causing Doppler-domain energy leakage.

Finally, before transmission, the AFDM signal must be prepended with a chirp-periodic prefix (CPP), which while structurally similar to the conventional cyclic prefix (CP) in OFDM, includes deterministic phase adjustments determined by the chirp parameters. With the CPP, AFDM satisfies the quasi-periodicity conditions and ensures the circular convolution compatibility to preserve the symbol orthogonality. At the receiver, demodulation is performed using a *matched* forward discrete affine Fourier transform (DAFT), which is configured with the same chirp parameters $c_1$ and $c_2$ used in the transmitter-side IDAFT, ensuring coherent symbol recovery.

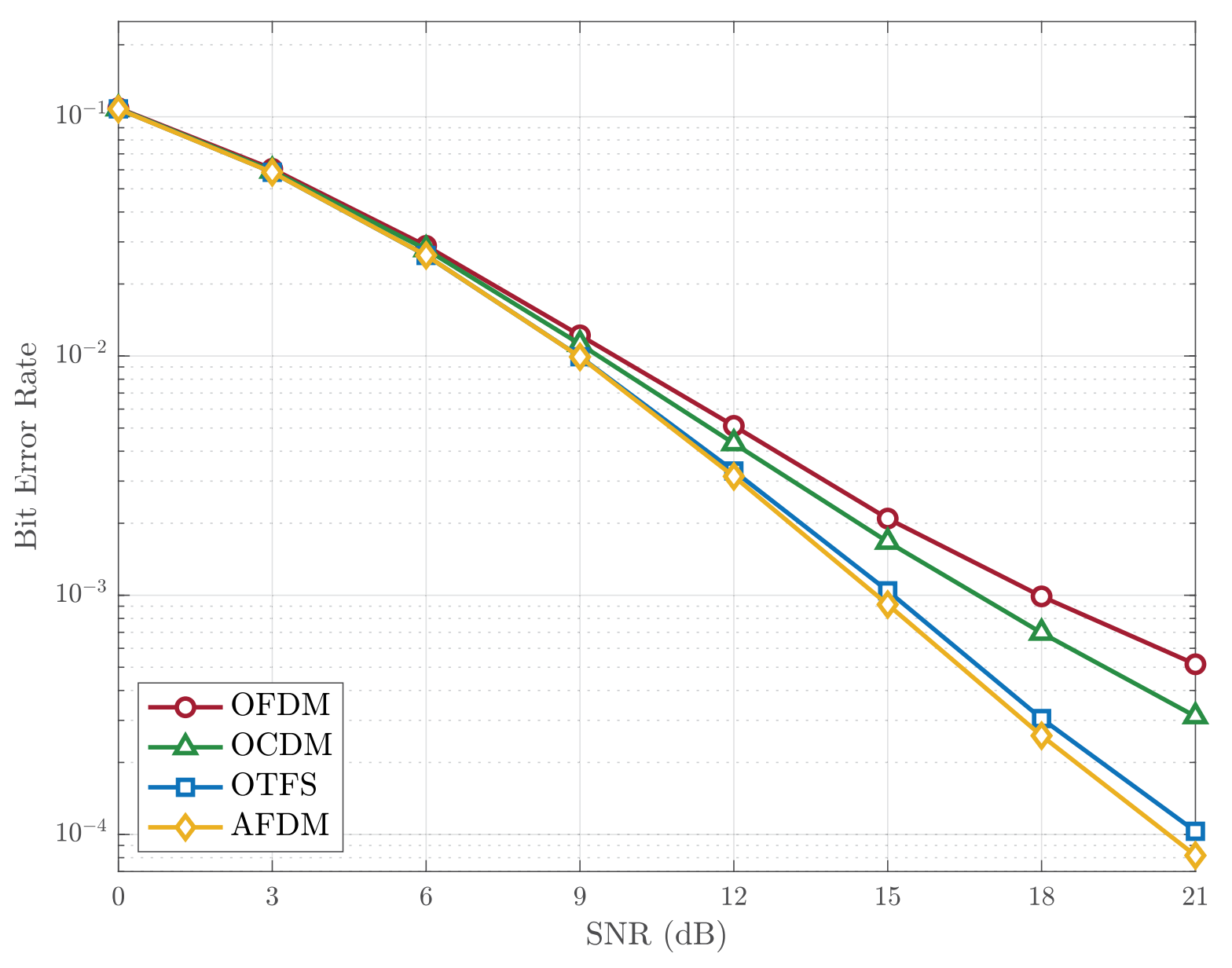

FIGURE 2. BER performance of OFDM, OCDM, OTFS, and AFDM over a doubly-dispersive channel with $P = 3$ paths. System parameters are: number of subcarriers $N = 64$, constellation cardinality $M = 4$, OTFS grid size $K \times L = 8 \times 8$, carrier frequency $f_c = 50$GHz, bandwidth $B = 150$MHz, AFDM guard parameter $\xi = 1$, AFDM pre-chirp parameter $c_2 = \frac{1}{2N}$.

For appropriately selected chirp parameters, whose optimal values can be obtained in closed-form based on the channel statistics (maximum unambiguous delay and Doppler shift), and a desired guard width, the effective channel matrix becomes a structured superposition of shifted band-diagonal components. Each such component corresponds to a distinct delay-Doppler path, and their orthogonality minimises ISI and ICI even under significant channel dispersion, further ensured by the selected guard width parameter.

### Affine Fourier Domain and the Evolution of 6G Waveforms

Chirp signals are well-known to have favorable properties under both time shifts (delays) and frequency shifts (Doppler), which intuitively suggests them to be well-matched to channels where such effects are jointly present, i.e., the doubly-dispersive channels. As described in the previous section, AFDM exploits this notion, and modulates the data symbols unto the orthogonal chirp subcarriers via the inverse discrete affine Fourier transform (IDAFT).

Trivially, the IDAFT originates from the affine Fourier transform (AFT), also referred to as the linear canonical transform (LCT), a family of unitary linear operators defined by four configurable parameters $(a, b, c, d)$, satisfying $ad–bc = 1$. By selecting appropriate values of the parameters $a$, $b$, $c$, and $d$, the AFT generalizes a wide range of well-known and fundamental transforms as its special cases.

For example, the AFT is reduced to the conventional Fourier transform (FT) with fixed parameters $(a, b, c, d) = (0, 1, -1, 0)$, while the fractional Fourier transform (FrFT) generalizes this by introducing a tunable angle parameter $\alpha \in [0; 2\pi)$ such that $(a, b, c, d) = (\cos\alpha, \sin\alpha, -\sin\alpha, \cos\alpha)$. On the other hand, the Fresnel transform (FnT), associated with free-space wave propagation and used in the modulation of OCDM, is obtained when $(a, b, c, d) = (1, z, 0, 1)$, where $z$ is a tunable parameter representing the propagation distance.

Furthermore, complex transforms including the Laplace transform (LT) can also be obtained, by setting the parameters as $(a, b, c, d) = \left(0, \sqrt{-1}, \sqrt{-1}, 0\right) = (0, j, j, 0)$, but such complex transforms are non-unitary and lack an orthonormal basis, rendering them unsuitable for energy-preserving signal modulation and orthogonal multiplexing for waveform design.

All in all, the AFT may be regarded as a generalized class of linear transforms that unifies and extends classical time-frequency operators, and AFDM as a natural generalization in the evolution of multicarrier modulation schemes, subsuming earlier schemes like OFDM and OCDM as special cases by introducing tunable chirp parameters $c_1$ and $c_2$, corresponding to the four AFT parameters $(a, b, c, d)$ [8].

## AFDM: Features and Applications

### Delay-Doppler Robustness and Diversity

One of the most critical requirements for 6G waveforms is the robustness to delay and Doppler dispersion - i.e., the ability to preserve orthogonality and full diversity in presence of rich multipath and Doppler shifts. This becomes more prominent in the envisioned high-frequency and dynamic environments, including urban V2X and SAGIN architectures.

The AFDM waveform stands out in this regard over other waveforms in the communications performance, as illlustrated in Fig. 2, offering inherent robustness to doubly-dispersive channels with arbitrary delay-Doppler statistics. The waveforms use the same total number of symbol resources, with OFDM, OCDM, and AFDM waveforms each employing $N$ time-frequency subcarriers, while OTFS operates over a delay-Doppler grid of size $K \times L$, where $K$ and $L$ denote the numbers of delay and Doppler bins, respectively, and $KL = N$.

Through the parametrizable chirp subcarriers, AFDM has been shown to guarantee full diversity without the need for precoding or equalization enhancements, by optimizing the first chirp parameter $c_1$, for which a closed-form solution is available [8]. This contrasts with conventional OFDM, where each symbol is confined to a single time-frequency subcarrier, resulting in a maximum diversity order of one in doubly-dispersive channels. While OTFS improves over the OFDM by spreading each symbol over the entire time-frequency grid via its delay-Doppler representation and transforms, uncoded OTFS has been shown to fall short of full diversity in the high-signal-to-noise ratio (SNR) regime unless mitigation strategies such as delay-Doppler precoding are applied [14].

Moreover, effective channel analysis has revealed that AFDM is inherently more robust fractional Doppler interference compared to other waveforms [4], [8], leveraging a tunable *guard parameter* to provide additional protection between each delay-Doppler tap.

## Pilot Design and Channel Estimation

Accurate and efficient channel estimation is fundamentally crucial for reliable wireless communications, where the consideration of pilot placement also is necessary in balancing estimation accuracy, spectral efficiency, and system latency.

In terms of pilotting efficiency, AFDM benefits from the inherent sparsity and full delay-Doppler representation enabled by its DAFT-domain formulation. A single well-placed pilot symbol, separated from data by only a small one-dimensional guarding determined by the maximum channel delay, can capture the entire delay-Doppler profile without data interference.

This stands in contrast to OFDM, which requires multiple pilot tones scattered across time and frequency, and OTFS, which requires larger two-dimensional guard symbols around the pilot to ensure data interference-free estimation. By eliminating the need for such extensive pilot resources, AFDM allows for tighter pilot-data packing, enhancing spectral efficiency and reducing overhead, which is particularly beneficial in latency- and bandwidth-constrained scenarios.

Building on this efficient pilot design, it has been shown that the full delay-Doppler channel in AFDM can be efficiently estimated using a single pilot symbol and low-complexity methods [8], thanks to the sparse effective channel structure composed of multiple band-diagonal components, unlike the block-shifted diagonal structure of OTFS or the interference-prone diagonal structure of OFDM [4], as illustrated in Fig. 1.

## CSI-Based Radar Parameter Estimation

Fundamentally, the full delay-Doppler representation of the doubly-dispersive channel is very closely linked to the radar parameters of environmental scatterers. Namely, each multipath component is associated with its delay and Doppler shift, which correspond to the bistatic range and velocity of the scatterer (or target).

This duality has motivated a wide range of methods for extracting such physical parameters from the estimated channel state information (CSI), or conversely to perform channel estimation directly via parameterized models that exploit the deterministic structure of the delay-Doppler channel, which has shown great success and efficiency [9], [11].

Such estimation techniques can be categorized into two main approaches: on-grid and off-grid methods. On-grid estimation involves discretizing the delay-Doppler space based on assumed maximum delay and Doppler values, followed by a search over this space, using either maximum likelihood estimation or compressive sensing-based sparse recovery. In contrast, off-grid approaches aim to circumvent discretization and perform estimation directly on the continuous delay-Doppler parameters, leveraging various techniques such as basis expansion methods and iterative algorithms.

## Conventional Radar Processing and Ambiguity Function

While CSI-based radar parameter extraction has become central to many modern communication-centric sensing frameworks, it is equally critical to consider the classical signal processing foundation of radar sensing: correlation-based matched filtering. This classical perspective is especially relevant for AFDM, whose underlying chirp subcarriers strongly resemble the gold standard of chirp structures historically used in radar systems. In this context, the AF is fundamental to quantify how effectively a probing waveform can resolve targets in the delay and Doppler dimensions, characterizing the resolution and sidelobe behavior of the waveforms.

Figure 3 illustrates the four waveforms using a fractionally interpolated (for a high-resolution representation of the continuous behavior) AF computed under unitary symbol modulation – i.e., all symbols are set to one. This highlights the intrinsic waveform- and transform-dependent characteristics in the delay-Doppler domain, independent of the behavior introduced by modulated random data symbols [10].

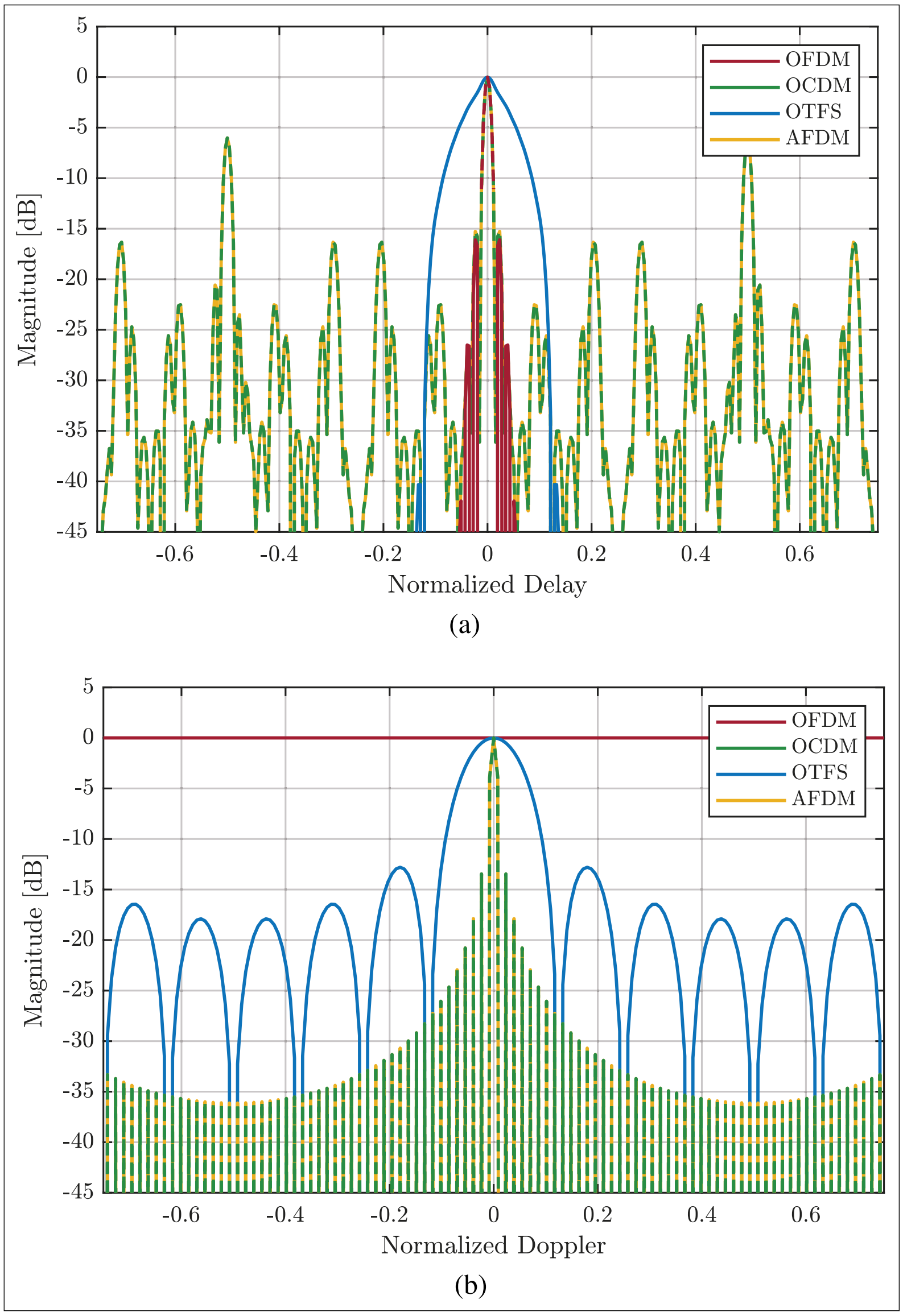

**FIGURE 3.** Interpolated (high-resolution) normalized ambiguity functions of OFDM, OCDM, OTFS, and AFDM in the delay-Doppler domain, with unitary sensing symbols, i.e., all symbols are set to one. System parameters are: number of subcarriers $N = 64$, OTFS delay-Doppler grid size $K \times L = 8 \times 8$, AFDM chirp parameters $c_1 = \frac{11}{2N}$, $c_2 = \frac{1}{2N}$. a) Zero-Doppler cut (delay response) of the waveform AFs. b) Zero-delay cut (Doppler response) of the waveform AFs.

The OFDM waveform exhibits a sharp peak along the delay axis but poor ambiguity characteristics in the Doppler domain, resulting in significant Doppler smearing and limited velocity resolution. In contrast, OTFS demonstrates improved joint localization in both delay and Doppler dimensions, albeit with broader mainlobes. Meanwhile, OCDM and AFDM achieve sharply concentrated, spike-like ambiguity functions in both delay and Doppler, reflecting their inherent chirp-based structure. In addition, AFDM is known to offer flexibility through the configuration of its chirp parameters, enabling delay-Doppler lattice alignment and controllable AF shaping.

### AFDM for Integrated Sensing and Communications (ISAC)

The above features of AFDM, particularly the excellent communications and radar ambiguity properties in doubly-dispersive channels, make it well-suited also for *communication-centric* ISAC, where the communication waveform itself is leveraged to extract high-resolution environmental information without requiring dedicated sensing signals, while also jointly supporting high-rate data transmission.

Indeed, recent works have shown that AFDM achieves accurate delay and Doppler parameter recovery with minimal sensing overhead but without sacrificing communication performance, leveraging embedded pilot modulation, subcarrier-level index modulation, and bilinear estimation methods for joint channel and data recovery [9], [11].

### Frequency-Domain Processing

Given the close structural relationship between AFDM and OFDM, certain configurations of the chirp parameters ($c_1$, $c_2$) and subcarrier length $N$, while still preserving full diversity and subcarrier orthogonality, admit closed-form frequency-domain representations of the AFDM waveform, enabling straightforward implementation and conventional analysis [15]. From this perspective, AFDM can be interpreted as a *precoded* OFDM system, where the precoder maps data symbols from the DAFT domain into the frequency domain through chirp-dependent rotations.

The trivial case for $c_1 = 0$, AFDM degenerates to standard OFDM up to a symbol-wise phase rotation and all OFDM-based frequency-domain techniques can be leveraged directly. However for $c_1 \neq 0$, the modulation introduces additional phase rotations before OFDM-like transformation. Still, the fundamental one-dimensional processing chain, where a complex symbol vector undergoes an DFT followed by another phase rotation, remains structurally identical to that of OFDM.

Accordingly, in non-doubly-dispersive channels (i.e., those exhibiting multipath delay but negligible Doppler), existing frequency-domain techniques developed for OFDM – such as pilot placement, channel estimation, and equalization – can be directly applied to AFDM. This strong structural alignment between the AFDM and OFDM processing chains implies that existing frequency-domain modules can be adapted with minimal modification for specific scenarios [13], facilitating a seamless evolution from 5G to 6G.

### Modulation Flexibility and Novel Applications

The tunable parameters of AFDM, $c_1$ and $c_2$, offer design flexibility absent in conventional waveforms like OFDM and OTFS. While $c_1$ must satisfy a minimal condition to ensure full diversity, $c_2$ remains largely unconstrained and can be tuned freely. Importantly, such configurations preserve the core benefits of AFDM, including diversity, pilot design simplicity, and favorable ambiguity characteristics.

This flexibility enables secondary functionalities such as index modulation via discrete $c_2$ sets or chirp permutations, and also enables physical-layer security features (see following section). These make AFDM especially suitable for systems requiring both high performance and embedded functionality within a unified, standard-compatible waveform.

### Inherent Physical-Layer Security

AFDM provides inherent physical-layer security through configurable the chirp parameters ($c_1$, $c_2$) and transform-domain methods such as chirp permutation without degrading performance. Legitimate users can treat these configurations as shared secret keys, while unauthorized receivers face infeasible brute-force search over a continuous parameter space, shown to be infeasible even under quantum-accelerated attacks. By contrast, OFDM employs fixed DFT-based subcarriers with no intrinsic secrecy, and while OTFS offers limited structural confidentiality via discrete grid size options, but the finite configuration space is constrained by channel diversity requirements. AFDM, with its broader and continuous parameter space, enables significantly stronger confidentiality.

## Standardization Considerations for 6G: Challenges, Opportunities, and Pathways

The preceding sections have provided a comprehensive overview of the AFDM waveform, outlining its key signal-theoretic features and practical advantages across multiple domains, including channel resilience, diversity, spectral properties, and computational efficiency. Building upon this foundation, this section explains how these technical features align with some emerging and current 6G requirements and design principles outlined by standardization bodies and academia, such as 3GPP [1], [12], International Telecommunication Union - Radiocommunication Sector (ITU-R) [2], and Institute of Electrical and Electronics Engineers (IEEE).

In this context, this section highlights the challenges and opportunities associated with integrating AFDM into the evolving 6G ecosystem, offering a roadmap that connects physical layer innovations and features of the waveform, with the practical deployment pathways, as summarized in Table 1, and also qualitatively compared in terms of the various candidate waveforms in Fig. 4.

### Backward Compatibility With the OFDM Standard

As depicted in Fig. 1, AFDM can be interpreted as a generalization of both OFDM and OCDM.

| Standardization Considerations for 6G (Section IV) | Orthogonal Chirp-Domain Subcarriers (II-A) | Generality of Affine Fourier Transform (II-B) | Delay-Doppler Robustness & Full Diversity (III-A) | Pilot Design & Channel Estimation (III-B) | CSI-based Radar Parameter Estimation (III-C) | Radar Processing & Ambiguity Function (III-D) | AFDM for Integrated Sensing and Commun. (III-E) | Frequency-Domain Processing (III-F) | Modulation Flexibility & Novel Applications (III-G) | Inherent Physical-Layer Security (III-H) |
|---|---|---|---|---|---|---|---|---|---|---|
| | **Fundamental Properties and Features of AFDM Waveform (Sections II-III)** | | | | | | | | | |
| **A. Backward Compatibility with the OFDM Standard** | | ✓ | | | | | | ✓ | | |
| **B. Multi-RAT Coexistence and Spectrum Sharing** | ✓ | ✓ | | | | | | ✓ | | ✓ |
| **C. PAPR and Hardware Efficiency** | ✓ | ✓ | | ✓ | | | | ✓ | ✓ | |
| **D. Ultra-Low Latency: Implementation Complexity** | | ✓ | | ✓ | ✓ | | | ✓ | | |
| **E. Standardized Parameter Configuration** | ✓ | | ✓ | ✓ | | | ✓ | | ✓ | ✓ |
| **F. Integration with Emerging Use Cases** | ✓ | | ✓ | ✓ | ✓ | ✓ | ✓ | | ✓ | ✓ |
| **G. Robustness to Practical Impairments** | ✓ | | ✓ | ✓ | ✓ | | ✓ | | | ✓ |
| **H. Testbeds, Trials, and Reference Designs** | ✓ | | ✓ | ✓ | ✓ | ✓ | ✓ | ✓ | ✓ | |

**TABLE 1.** Mapping of AFDM properties and features (see "Fundamentals and Key Properties" and "AFDM: Features and Applications" sections) to relevant 6G standardization considerations (see "Standardization Considerations for 6G: Challenges, Opportunities, and Pathways" section).

Furthermore, as identified in the "Frequency-Domain Processing" section, since frequency-domain processing techniques developed for OFDM are directly applicable to AFDM, the transition to AFDM can be achieved with minimal cost by leveraging existing OFDM infrastructure.

In other words, upgrading an OFDM system to AFDM merely involves the inclusion of a precoding block at the transmitter and a corresponding decoding block at the receiver. This modification is particularly cost-effective in software-defined radio platforms, and enables the seamless reuse of a wide range of components and protocols already specified in standards such as those from 3GPP and IEEE. As a result, AFDM supports a cost-efficient migration path from 5G to 6G, facilitating incremental upgrades and hybrid deployments during early stages of 6G adoption.

While similar compatibility arguments can also be made for OTFS, AFDM exhibits greater flexibility and robustness, enhancing its backward compatibility with OFDM. As previously discussed, AFDM achieves full diversity for any number of subcarriers $N$, depending solely on the chirp parameters $c_1$ and $c_2$. In contrast, full diversity in OTFS is only possible for certain divisors of $N$, a constraint that becomes particularly limiting in narrowband applications.

This distinction is especially relevant in use cases such as narrowband (NB)-IoT, where the number of subcarriers may be as small as $N = 12$, corresponding to a single resource block defined in the 3GPP specification. In such scenarios, only AFDM can simultaneously ensure backward compatibility with OFDM and robustness to doubly dispersive channels, whereas OTFS falls short in this regard.

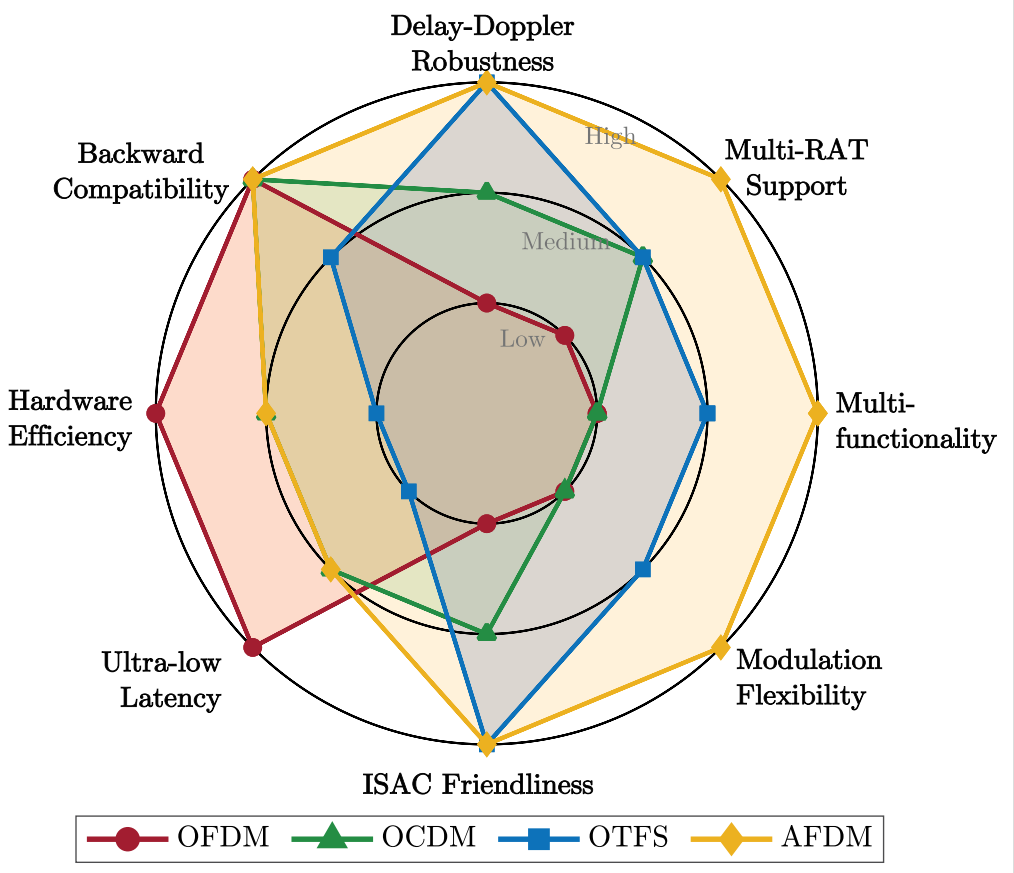

**FIGURE 4.** Qualitative comparison of the features and standardization-related properties of various waveforms: OFDM, OCDM, OTFS, and AFDM, which illustrates that AFDM is the best all-rounder waveform current under consideration for ISAC-enabling 6G systems, without significant sacrifice compared to OTFS and OFDM in terms of their specific advantages in hardware efficiency and low latency.

### Multi-RAT Coexistence and Spectrum Sharing

As the wireless landscape becomes increasingly heterogeneous, waveform compatibility across multiple radio access technology (RAT) (multi-RAT) is essential for enabling seamless coexistence and efficient spectrum utilization. The previously elaborated compatibility of AFDM with OFDM, provides a strong foundation for cost-effective migration from 5G to 6G, specifically in facilitating hybrid deployments where multiple waveforms share spectrum or infrastructure.

In particular, the spectral and structural similarity between AFDM and OFDM allows for flexible spectrum partitioning and orthogonal resource allocation, enabling both waveforms to operate side-by-side with minimal interference - as shown by AFDM reducing to the OFDM with appropriate chirp parameters. This is especially relevant for legacy support, where incumbent 5G services must coexist with new 6G features during the early stages of deployment. These attributes support the vision of future networks where spectrum resources are flexibly and efficiently allocated across diverse services and technologies across generations, in alignment with evolving ITU-R and 3GPP spectrum sharing frameworks.

### PAPR and Hardware Efficiency

Hardware efficiency is a key consideration for waveform adoption in future wireless standards. PAPR, in particular, remains a critical issue, as high values reduce power amplifier efficiency and raise hardware costs, especially in mobile and battery-powered 6G devices. Therefore, while AFDM offers strong delay-Doppler performance, maintaining favorable peak-to-average power ratio (PAPR) characteristics is still essential for practical deployment.

To address this issue, techniques such as chirp parameter tuning and DAFT-domain spreading have been proposed specifically for AFDM, providing design flexibility without degrading performance in doubly-dispersive channels. Additionally, the compatibility with OFDM systems also enables the reuse of established PAPR reduction methods, including tone reservation, windowing, and pulse shaping.

Furthermore, as will be elaborated in the following, compared to waveforms like OTFS, AFDM has been shown to achieve a lower-complexity modulator structure, reducing hardware overhead and improving energy efficiency across diverse 6G use cases, from high-speed access to massive machine-type communications.

### Ultra-Low Latency: Implementation Complexity

The practical adoption of any new waveform hinges on its ability to support real-time processing under stringent latency and complexity constraints, especially in 6G systems that demand extremely low latency and high throughput.

Although AFDM relies on the advanced DAFT, it was shown - as discussed in previous sections - to be implemented efficiently using structured fast transform algorithms, remaining tractable for even larger systems increasingly expected in 6G networks.

In addition to low-complexity modulation and demodulation, AFDM also benefits significantly from the sparse and structured effective channel response, enabling simplified equalization and efficient pilot design. Compared to OTFS, which requires a two-dimensional piloting guard for effective channel estimation and typically more complex signal processing for channel reconstruction, AFDM offers a more streamlined receiver architecture. These advantages position AFDM as a waveform that can increasing meet real-time requirements across a wide range of devices, from high-performance base stations to resource-constrained user terminals.

### Standardized Parameter Configuration

As discussed, one of the defining feature of AFDM is its flexible parametrization of continuous-valued chirp frequencies ($c_1$, $c_2$), enabling adaptation to channel conditions, system requirements, and additional applications. However, this flexibility will also necessitate standardized procedures for parameter sharing and signaling. Unlike OFDM, where subcarrier spacing and symbol duration are fixed by the numerology, AFDM supports a broader configuration space. Therefore, to ensure interoperability and consistent performance across vendors and deployments, standards must define permissible values or ranges for the chirp parameters, potentially based on link type, mobility profile, or application class.

Novel signaling mechanisms must also be considered to convey the chirp configurations in both uplink and downlink, covering various components such as initial access, adaptive reconfiguration during handover, and multi-cell coordination. Well-defined parameter profiles and negotiation protocols will be essential to minimize ambiguity, reduce signalling overhead, and ensure consistent system behavior in deployment.

### Integration With Emerging Use Cases

The adoption of any new waveform in 6G depends on its ability to reliably and efficiently support emerging use cases and service requirements. AFDM demonstrates strong potential in this regard – especially under the ISAC framework – owing to its inherent flexibility, delay-Doppler robustness, and processing efficiency, well-aligned with key 6G paradigms.

For instance, AFDM is particularly well-suited for high-mobility scenarios such as V2X communications and SAGIN architectures involving low Earth orbit (LEO) constellations and autonomous aerial vehicles (AAVs), due to its resilience to Doppler shifts and delay spreads, supporting the vision of key verticals like automotive and future vehicular services.

In addition, the flexible design of AFDM supports advanced multifunctional applications, as shown by the previous discussions on the excellent integration of AFDM with ISAC framework, but also a plethora of other next-generation technologies such as semantic communications and over-the-air computing, both emerging enablers of future network intelligence.

Together, these capabilities position AFDM not only as an efficient transmission waveform, but as a flexible platform for addressing the diverse functional requirements of 6G.

### Robustness to Practical Impairments

Robustness to real-world impairments remains a key requirement for waveform standardization,

particularly for deployment in diverse hardware environments and under non-ideal channel conditions. Such impairments include carrier frequency offset (CFO), non-white noise, amplifier nonlinearities, and quantization effects, all of which are known to degrade multicarrier waveforms like OFDM.

As evaluated, the AFDM exhibits natural resilience to these challenges, where the spreading in the DAFT domain distributes energy across time and frequency, mitigating sensitivity to CFO and providing enhanced robustness to colored noise. These properties position AFDM as a robust and scalable candidate for practical 6G scenarios where performance must be preserved despite front-end imperfections.

### Testbeds, Trials, and Reference Designs

Finally, the transition from academic exploration to standardization necessitates rigorous validation through hardware testbeds, field trials, and open reference implementations. While AFDM has demonstrated strong theoretical performance across key metrics such as diversity gain, delay-Doppler resilience, and reduced ambiguity, its real-world viability remains to be fully established.

To this end, prototyping efforts based on software-defined radios and open-source platforms are essential to characterize implementation complexity, system-level performance, and compatibility with existing communication stacks. Such platforms can enable early experimentation with channel estimation, synchronization, and multi-user access techniques tailored to AFDM. These practical efforts will be instrumental in demonstrating the deployability of AFDM at scale and ensuring that its performance gains translate effectively into real-world environments, an essential step toward 6G adoption.

## Conclusion and Outlook

AFDM has emerged as a compelling candidate for the physical layer machinery of 6G systems, uniting robustness to delay-Doppler dispersion, full diversity, integrated sensing compatibility, and waveform-level flexibility within a single framework. The parametrized structure enables not only high communication performance but also opens new design frontiers for multi-functional operation.

Meanwhile, its backward compatibility, along with its readiness for frequency-domain processing, makes it particularly attractive for adoption in 6G, allowing cost-effective migration and coexistence in multi-RAT environments. Yet, the path to standardization remains open and requires focused development and discussions including bench-marking and link-and-system level demonstrations to further confirm our analysis.

All in all, with growing interest from both academia and industry, AFDM stands poised to help define the waveform landscape of the next generation of wireless standardization.

## Biographies

Hyeon Seok Rou (Member, IEEE) (hrou@constructor.university) received the B.Sc. degree in electrical and computer engineering and the Ph.D. degree in electrical engineering from Constructor University Bremen (formerly Jacobs University Bremen), Germany, in 2021 and 2024, respectively. He was a Visiting Researcher with the Intelligent Communications Laboratory (ICL!), KAIST, in 2023. Since 2025, he has been a Post-Doctoral Research Fellow and a Lecturer with the School of Computer Science and Engineering, Constructor University Bremen. His research interests include next-generation wireless systems, affine frequency division multiplexing (AFDM), integrated sensing and communications (ISAC), high-mobility communications, multi-dimensional sparse modulation (index modulation), over-the-air computing (AirComp), and quantum-accelerated optimization. He received Korea Institute of Science and Technology Europe Research Scholarship Award in 2022. He was German Delegate to the Young Professionals Forum (YPF) at the 3rd World Congress of Korean Scientists and Engineers, South Korea, in 2025. He has been serving as the Chapter Representative for The Korean Scientists and Engineers Association in the FRG (VeKNI e.V.) since 2021. He also contributes to the IEEE Standards Association P3383 Working Group on Performance Metrics for Integrated Sensing and Communication.

Kuranage Roche Rayan Ranasinghe (Graduate Student Member, IEEE) (kranasinghe@constructor.university) received the B.Sc. degree in electrical and computer engineering and the B.Sc. degree in robotics and intelligent systems (with a minor in physics) from Constructor University (formerly Jacobs University, Bremen) in 2023 and 2024, respectively, where he is currently pursuing the Ph.D. degree in electrical engineering. He was a Visiting Researcher at ETH Zürich, Switzerland, and the KTH Royal Institute of Technology, Sweden, in 2025. Within the fields of wireless communications and signal processing, his research interests include integrated sensing, communications and computing (ISCC), compressed sensing, Bayesian inference, next-generation metasurface technologies, and optimization theory. He received the Best Paper Award from the International Conference on Computing, Networking and Communications (ICNC) in 2025. He also contributes to the IEEE Standards Association P3383 Working Group on Performance Metrics for Integrated Sensing and Communication.

Vincent Savaux (Senior Member, IEEE) (vincent.savaux@b-com.com) received the Engineering degree from the High School of Engineering, ECAM Rennes, the M.Sc. degree from the University of Rennes 1, France, in 2010, and the Ph.D. degree in telecommunications from the École Supérieure d'Electricité-Supélec in 2013. From 2014 to 2015, he was a Post-Doctoral Researcher with the Signal, Communication, and Embedded Electronics (SCEE) Research Group, CentraleSupélec, Campus de Rennes, France. Since 2016, he has been a Research Engineer with the Institute of Research and Technology b-com, Rennes, France. His research activities mainly focus on signal processing for wireless systems, including the IoT technologies, MIMO systems, and new waveforms for next generations of communications systems. He received the Exemplary Reviewer 2014 Appreciation from IEEE Wireless Communications Letters and IEEE Communications Letters.

Giuseppe Thadeu Freitas de Abreu (Senior Member, IEEE) (gabreu@constructor.university) received the B.Eng. degree in electrical engineering and the Specialization Latu Sensu degree in telecommunications engineering from the Universidade Federal da Bahia (UFBa), Salvador, Bahia, Brazil, in 1996 and 1997, respectively, and the M.Eng. and D.Eng. degrees in physics, electrical, and computer engineering from Yokohama National University, Japan, in March 2001 and March 2004, respectively. He was a Post-Doctoral Fellow and later an Adjunct Professor (Docent) in statistical signal processing and communications theory with the Department of Electrical and Information Engineering, University of Oulu, Finland, from 2004 to 2006 and from 2006 to 2011, respectively. Since 2011, he has been a Professor of electrical engineering with Constructor University Bremen (formerly Jacobs University, Bremen, Germany). From April 2015 to August 2018, he was a Full Tenured Professor with the Department of Computer and Electrical Engineering, Ritsumeikan University, Japan. He has been a Visiting Scholar with Rice University in 2008, Stanford University in 2009, Keio University in 2010, the University of Bologna in 2012 and 2014, and the KTH Royal Institute of Technology in 2025. He has also been a Regular Visiting Professor with the University of Electro-Communications, Japan, since 2013. His research interests include a wide range of topics on wireless communications and signal processing, including communications theory, estimation theory, optimization theory, statistical modeling, metasurfaces and parameterizable electromagnetic structures, waveform design, integrated sensing and communications, over-the-air computing, wireless localization, cognitive radio, wireless security, MIMO systems, ultrawideband and millimeter wave communications, full-duplex and cognitive radio, compressive sensing, energy harvesting networks, random networks, connected vehicles networks, and many more. He received the Uenohara Award from Tokyo University in 2000 for his master's thesis work. He was a co-recipient of awards from various international IEEE conferences, including the Asilomar in 2009, 2012, and 2014, WPNC in 2012 and 2015, ISWCS in 2016, WPMC in 2023, and ICNC in 2025. He was also a co-recipient of the Best Journal Paper Award from Japanese Chapter of the IEEE Signal Processing Society in 2023 and was awarded various prestigious Fellowships from the Heiwa Nakajima Foundation, the JSPS, and the NICT (twice), in 2010, 2013, 2015, and 2018, respectively. He served as an Editor for IEEE Transactions on Wireless Communications from 2009 to 2014, IEEE Transactions on Communications from 2014 to 2017, and IEEE Communication Letters from 2021 to 2024, and as an Executive Editor for IEEE Transactions on Wireless Communications from 2017 to 2021. He is currently serving as an Editor for IEEE Signal Processing Letters and IEEE Open Journal of the Communications Society.

David González G. (Senior Member, IEEE) (david.gonzalez.g@ieee.org) received the master's degree in mobile communications and the Ph.D. degree in signal theory and communications from the Universitat Politècnica de Catalunya (UPC), Spain. From 2014 to 2017, he was a Post-Doctoral Research Fellow with Aalto University, Finland. From 2017 to 2018, he was a Research Engineer with the Panasonic Research and Development Center, Germany. Since 2018, he has been with AUMOVIO Germany GmbH (formerly Continental Automotive Technologies, Germany), where he leads and contributes to research projects on wireless and vehicular communications (V2X), integrated sensing and communications (ISAC), and automotive connectivity for 5G-Advanced and 6G systems. He actively participates in pre-standardization and standardization activities as a delegate to 3GPP RAN1, the 5G Automotive Association (5GAA), and ETSI ISG ISAC.

Christos Masouros (Fellow, IEEE) (c.masouros@ucl.ac.uk) received the Diploma degree in electrical and computer engineering from the University of Patras, Greece, in 2004, and the M.Sc. and Ph.D. degrees in electrical and electronic engineering from the University of Manchester, U.K., in 2006 and 2009, respectively. In 2008, he was a Research Intern at Philips Research Laboratories, U.K., working on the LTE standards. From 2009 to 2010, he was a Research Associate with The University of Manchester, U.K., and from 2010 to 2012, a Research Fellow with Queen's University Belfast. In 2012, he joined University College London as a Lecturer. He has held a Royal Academy of Engineering Research Fellowship from 2011 to 2016. Since 2019, he has been a Full Professor of signal processing and wireless communications with the Information and Communication Engineering Research Group, Department of Electrical and Electronic Engineering, and affiliated with the Institute for Communications and Connected Systems, University College London. His research interests include wireless communications and signal processing with a particular focus on green communications, large-scale antenna systems, integrated sensing and communications, interference mitigation techniques for MIMO, and multicarrier communications. From 2018 to 2022, he was the Project Coordinator of the €4.2m EU H2020 ITN Project PAINLESS, involving 12 EU partner universities and industries, toward energy-autonomous networks. From 2024 to 2028, he will be the Scientific Coordinator of the €2.7m EU H2020 DN Project ISLANDS, involving 19 EU partner universities and industries, toward next-generation vehicular networks. He is a member of the IEEE Standards Association Working Group on ISAC performance metrics, a Founding Member of the ETSI ISG on ISAC, and a fellow of the Institution of Engineering and Technology (IET), the Artificial Intelligence Industry Alliance (AIIA), and Asia-Pacific Artificial Intelligence Association (AAIA). He was a recipient of the 2024 IEEE SPS Best Paper Award, the 2024 IEEE SPS Donald G. Fink Overview Paper Award, the 2023 IEEE ComSoc Stephen O. Rice Prize, and the Best Paper Awards from the IEEE GlobeCom 2015 and IEEE WCNC 2019 conferences, and a co-recipient of the 2021 IEEE SPS Young Author Best Paper Award. He is an IEEE ComSoc Distinguished Lecturer from 2024 to 2025, and his work on ISAC has been featured in the World Economic Forum's report on the top 10 emerging technologies. He is also a Founding Member and the Vice-Chair of the IEEE Emerging Technology Initiative on Integrated Sensing and Communications (SAC), and the Chair of the IEEE SPS ISAC Technical Working Group and the IEEE Green Communications and Computing Technical Committee, Special Interest Group on Green ISAC. He is the TPC Chair of the IEEE ICC 2024 Selected Areas in Communications (SAC) Track on ISAC, the Chair of the IEEE PIMRC2024 Track 1 on PHY and Fundamentals and the "Integrated Imaging and Communications" stream in IEEE CISA 2024, and the TPC Co-Chair of IEEE VTC 2025. He has been recognized as an Exemplary Editor of IEEE Communications Letters and an Exemplary Reviewer of IEEE Transactions on Communications. He is an Area Editor of IEEE Transactions on Wireless Communications and the Editor-at-Large of IEEE Open Journal of the Communications Society. He has been an Editor of IEEE Transactions on Communications, IEEE Transactions on Wireless Communications, and IEEE Open Journal of Signal Processing, an Associate Editor of IEEE Communications Letters, and a Guest Editor of a number of IEEE Journal on Selected Topics in Signal Processing issues.